\documentclass{article}
\usepackage{spconf,amsmath,graphicx}
\usepackage{kotex}
\usepackage{subcaption}
\usepackage{caption}
\usepackage{multirow}
\usepackage{makecell}
\usepackage{amssymb}
\usepackage{xcolor}
\usepackage{courier}

\DeclareMathOperator*{\argmin}{arg\,min}

\title{Joint unsupervised and supervised learning  for \\ context-aware language identification}

\name{Jinseok Park, Hyung Yong Kim, Jihwan Park, Byeong-Yeol Kim, Shukjae Choi, Yunkyu Lim}
\address{42dot Inc., Seoul, Republic of Korea \\
\footnotesize{\texttt{\{jinseok.park, hyungyong.kim, jihwan.park, byeongyeol.kim, shukjae.choi, yunkyu.lim\}@42dot.ai}}}

\begin{document}
\maketitle
\begin{abstract}
Language identification (LID) recognizes the language of a spoken utterance automatically. 
According to recent studies, LID models trained with an automatic speech recognition (ASR) task perform better than those trained with a LID task only. 
However, we need additional text labels to train the model to recognize speech, and acquiring the text labels is a cost high.
In order to overcome this problem, we propose context-aware language identification using a combination of unsupervised and supervised learning without any text labels. The proposed method learns the context of speech through masked language modeling (MLM) loss and simultaneously trains to determine the language of the utterance with supervised learning loss. The proposed joint learning was found to reduce the error rate by 15.6\% compared to the same structure model trained by supervised-only learning on a subset of the VoxLingua107 dataset consisting of sub-three-second utterances in 11 languages.
\end{abstract}


\section{Introduction}
Language identification (LID), the process of automatically recognizing the language of a spoken utterance, is mainly used as a precursor stage of multi-lingual speech recognition and translation. 
Various features are available to distinguish languages, from low-level acoustic and prosodic features (tone, stress, duration, rhythm) to high-level features such as morphology and sentence syntax~\cite{li2013spoken}, and many LID methods have been proposed with these features~\cite{zissman1996comparison, zissman2001automatic}.

Over the last decade, many language identification methods with deep neural networks (DNNs) have been proposed~\cite{lopez2014automatic, gonzalez2015frame, wang2022attentive}. Those approaches including learnable feature extractors achieved better performance than feature-based LID in challenging conditions such as utterances shorter than three seconds.
Furthermore, according to recent studies~\cite{tang2017phonetic, duroselle2021modeling, punjabi2021joint}, LID models trained with an automatic speech recognition (ASR) task perform better than those trained with a language identification task only. This implies that recognizing the context of utterances helps to identify the language spoken. 

Although ASR helps improve LID performance, we need additional text labels to train the model for the ASR task. While language labels for each utterance are relatively easy to obtain, acquiring text labels for utterances is cost high. This follows because people roughly recognize a spoken language even if they do not know it well, whereas text labels of speech must be written by a native or expert who knows the language well.

In this paper, we propose joint unsupervised and supervised learning for context-aware language identification that does not need additional text labels. The proposed method learns the speech context through masked language modeling (MLM) loss and simultaneously trains to determine the language of the utterance with supervised learning loss. To show the effect of the proposed joint learning in wild datasets, we collected 11,000 hours of speech data in 11 languages from YouTube for training and used a subset of the VoxLingua107~\cite{valk2021voxlingua107} consisting of sub-three-second utterances. Our experimental results show that the proposed joint learning approach reduced the error rate by 15.6\% compared to the same structure model trained on a supervised-only basis.

\section{Related Work}
\vspace{2mm}\noindent\textbf{LID with ASR:}
In~\cite{tang2017phonetic}, Tang et al. remarked that phonetic information has largely been overlooked in the previous DNN-based LID and showed that the phonetic information acquired by training the ASR task improves the performance of LID.
Duroselle et al.~\cite{duroselle2021modeling} conducted experiments in various combinations to measure how ASR training affects LID performance. In most cases, LID models trained with ASR generally performed better than models trained to identify languages only.
In~\cite{punjabi2021joint}, Punjabi et al. showed that when their model trained jointly for ASR and LID tasks on English and Hindi datasets, performance on both two tasks improved.
Although it has been experimentally proven that ASR learning helps to improve LID performance in various studies like this, a limitation is that a text label corresponding to each utterance is required to incorporate ASR learning when learning the LID model.

\vspace{2mm}\noindent\textbf{BEST-RQ:}
Chiu et al.~\cite{chiu2022self} proposed BERT-based speech pre-training with a random-projection quantizer (BEST-RQ). They generated discrete labels from a speech signal using the random-projection quantizer and demonstrated that the network, as trained using an MLM loss to predict the discrete labels generated, showed good performance in ASR. This implies that the model learned the speech context by BEST-RQ, and we use the BEST-RQ as our unsupervised learning of LID.

\vspace{2mm}\noindent\textbf{JUST:}
Bai et al.~\cite{bai2022joint} proposed joint unsupervised and supervised training (JUST) for multilingual ASR. Since JUST trains the model by jointly calculating unsupervised loss and supervised loss, the model will be optimized to a single loss, in contrast to the conventional self-supervised models which are optimized for each loss in different steps.
JUST also showed good performance when training the existing multi-lingual ASR data only without additional data.
We use this joint learning concept in this paper for that the LID models recognize the speech context by end-to-end learning without additional training data.

\section{Two learning methods}
\subsection{Supervised learning}
In order to train the LID model by supervised learning, we first prepare pair data of input speech and language label $\mathbf{y}$. 
We then extract the features $\{\mathbf{x}_t\}_{t=1}^T$ for $T$ time steps from the input speech signal.
The features are masked along the time axis with the masking probability $\rho_m$ for masking size $m$. We skip the masking step if the masking size is zero or in the inference process.

After calculating the represented vectors by passing features to the encoder layers, the language probability vector
$\mathbf{p} \in \mathbb{R}^N$ 
is calculated by a pooling layer, a linear classifier, and a softmax layer. Linear classifiers have output size $N$ signifying the number of language classes. Finally, cross-entropy loss for supervised learning is calculated with language probability  $\mathbf{p}$ and language label $\mathbf{y}$, as

\begin{equation}
\mathcal{L}_{s} = -\sum_{i=1}^N y_i \log p_i.
\end{equation}

\begin{figure}[t]
\begin{center}
\includegraphics[width=0.92\linewidth]{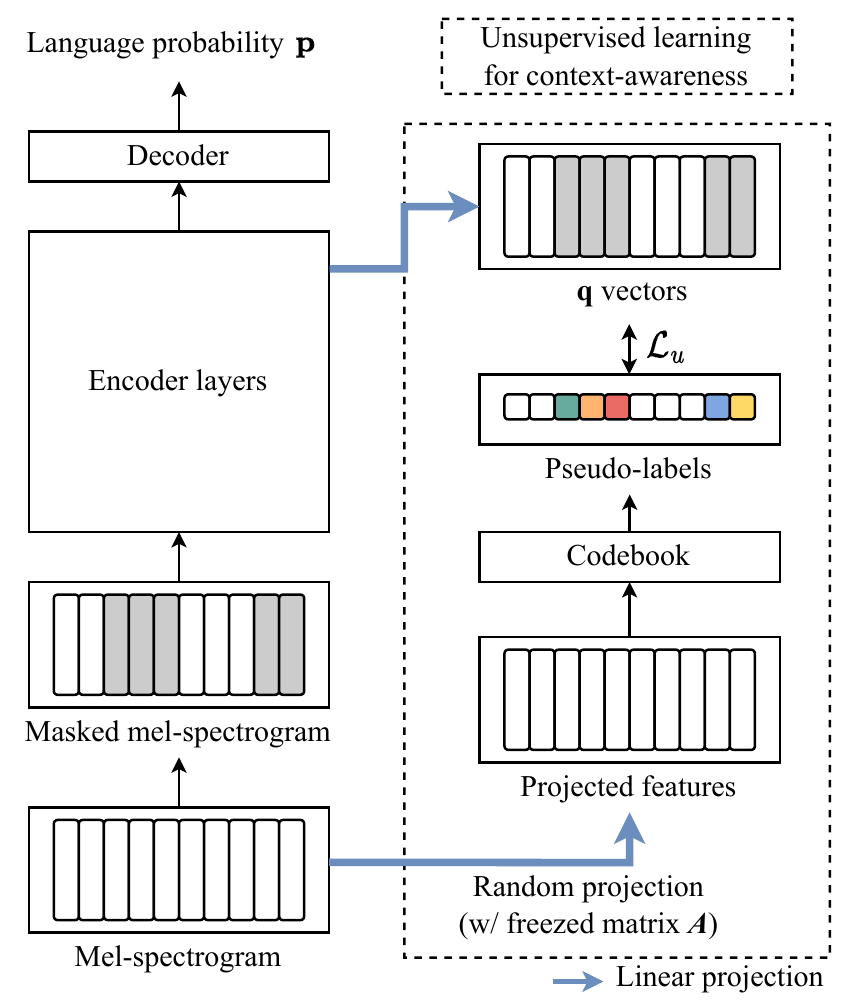} 
\end{center}
\caption{The overall learning process. The left side represents the normal language identification model (baseline) and the right side represents the unsupervised learning process (the sub-sampling concept is omitted).} %Figure 의 caption
\label{fig-architecture}
\end{figure}

\subsection{Joint learning}
We apply BEST-RQ~\cite{chiu2022self} as unsupervised learning to help the model catch the context from the input speech signal. First, we generate codebook vectors $ \{\mathbf{c}_j\}_{j=1}^M$
and a random projection matrix $A\in \mathbb{R}^{D\times F}$ with Xavier uniform initialization~\cite{glorot2010understanding} before starting training; here, $D$ is the dimension of the codebook vector, $F$ is the frequency dimension of the input feature, and $M$ is the number of codebook vectors. We then normalize the codebook vectors and fix $ \{\mathbf{c}_j\}_{j=1}^M$ and $A$.

During training, we generate pseudo-labels $\mathbf{z}_t$ for each time step $t$ by comparing distances among the $ \{\mathbf{c}_j\}_{j=1}^M$ to the normalized projected feature calculated by normalizing and multiplying $A$ and $\mathbf{x}_t$:

\begin{equation}
\mathbf{z}_t  \leftarrow \argmin_j | Norm(A\mathbf{x}_t) - \mathbf{c}_j |.
\end{equation}

We generate vectors $\{\mathbf{q}_t\}_{t=1}^T$ from the intermediate vectors of an encoder layer by linear projection and softmax. 
The model will be trained to classify the pseudo-label of the masked position $\mathcal{T}$, and the unsupervised loss is as follows: 

\begin{equation}
\mathcal{L}_{u} = - \frac{1}{|\mathcal{T}|} \sum_{t\in \mathcal{T}} \sum_{j=1}^M z_{t,j} \log q_{t,j}.
\end{equation}

In the above process, if the encoder includes a sub-sampling layer, the input features are stacked to the same degree as the sub-sampling factor. The random projection matrix is also generated to have the same dimension as the stacked input feature. 

Joint learning uses the losses of both supervised and unsupervised learning at the same time, and the final loss $\mathcal{L}$ is calculated as

\begin{equation}
\mathcal{L} = (1-\lambda) \mathcal{L}_{s} + \lambda \mathcal{L}_{u},
\end{equation}
where $\lambda$ has a weighting value between $[0,1]$ and is set to 0.5 in our experiments.
Fig. \ref{fig-architecture} shows the overall learning process.

\section{Experiments}
\subsection{Training setup}
\vspace{2mm}\noindent\textbf{Features:} All speech signals were converted into the log-mel spectrogram. The features were extracted with window size 400, hop size 160, fast Fourier transform size 512, and 80 filters. The sample rate was 16,000. 

\vspace{2mm}\noindent\textbf{Architecture:} 
We used the Conformer(S)~\cite{gulati2020conformer} encoder including the sub-sampling layer. All parameters are the same as for the original Conformer(S): the number of encoder layers is 16, the encoder dimension is 144, and the number of attention heads is 4. For the decoder, we used an average pooling layer and a linear classifier. 

\vspace{2mm}\noindent\textbf{Unsupervised learning parameters:} 
We set the codebook vector dimension $D$ as 16 and the number of codebook vectors $M$ as 64, 256, and 1024. 
The $\mathbf{q}$ vectors were generated from the output of the 15th encoder layer.
The masking probability $\rho_m$ was set differently depending on the masking size to make the average of the total masking region 35\%. Masking regions can be overlapped. We conducted the experiment by dividing the masking size into six steps from 80 ms to 480 ms.

\vspace{2mm}\noindent\textbf{Datasets and training parameters:} 
Training and evaluation were conducted for 11 languages\footnote{English (en), French (fr), German (de), Hindi (hi), Indonesian (id), Italian (it), Japanese (ja), Korean (ko), Mandarin Chinese (zh), Spanish (es), and Vietnamese (vi).}. To train with a large amount of data for each language class, we collected data from YouTube manually. 
We listed over a hundred channels for each language, including news, documentary, talk show, beauty channels, etc., and downloaded audio data from each channel.
We used our voice activity detection to extract speech segments from the downloaded audio data, acquiring 1,000 hours of utterances for each language for a total of 11,000 hours. The total number of segments is 4.8 million, with an average segment length of 8.16 seconds. For training, the Adam optimizer~\cite{kingma2014adam} was used with the peak learning rate set to 0.001 and warm-up set to 5,000 steps. All input signals were cropped to three seconds at a random position during training, and the batch size was 256.

\begin{figure}[t]
\centering
\begin{subfigure}{1\linewidth}
  \includegraphics[width=1\linewidth]{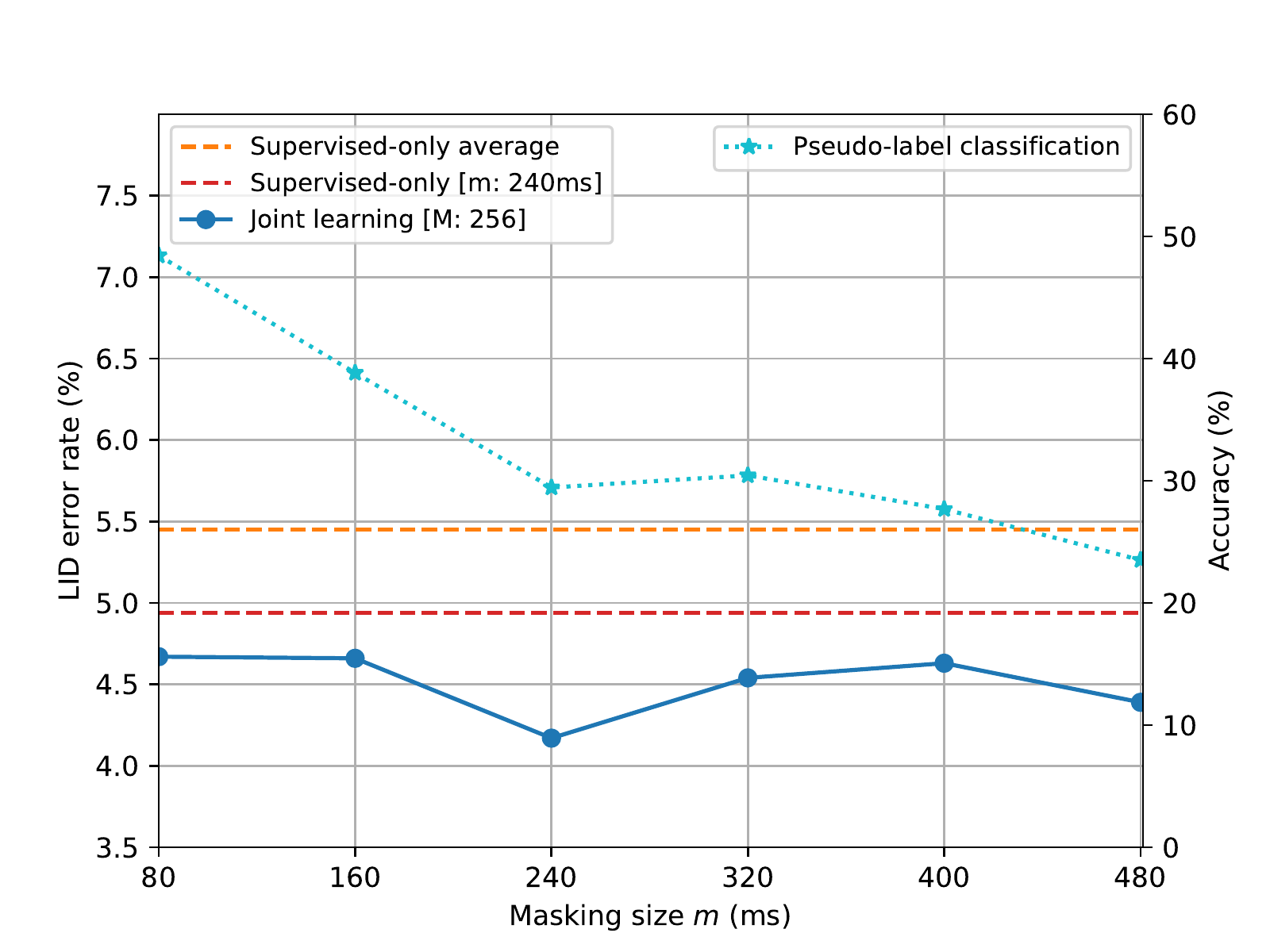} %TNR 로 변경하기. (FPR는 5~10% 정도로 설정)
\end{subfigure}
\caption{The results for masking size $m$. The left axis represents the error rates (\%) of LID; the right axis represents the pseudo-label classification accuracy (\%).}
\label{fig-ablation}
\end{figure}

\begin{table} [t]
\caption{LID classification error rate (\%) on VoxLingua-T11. The top three rows show the results according to the model structure and the bottom four rows show the results according to learning methods.} 
\begin{center}
\begin{tabular}{l | c } 
\Xhline{2\arrayrulewidth}
Method &  VoxLingua107-T11(3s)  \\ \hline  
\small{w/o masking (\#parameters)} & \\
ECAPA-TDNN (15.7M)~\cite{desplanques2020ecapa}  & 6.54\% \\ 
ATP-Conformer (9.5M)~\cite{wang2022attentive}  & 6.08\% \\ 
Supervised-only (10.3M)   & 5.80\% \\ 
\hline
\small{w/ masking [$M / m$]} & \\
Supervised-only [- / 240 ms]  & 4.94\% \\  
Joint learning [64 / 240 ms]  & 4.23\% \\ 
Joint learning [256 / 240 ms]  & \textbf{4.17\%} \\
Joint learning [1024 / 240 ms]  & 4.33\% \\
\Xhline{2\arrayrulewidth}
\end{tabular}
\end{center}
\label{table-multi-1} 
\end{table}

\begin{figure*}[t]
\centering
\begin{subfigure}{0.45\linewidth}
  \includegraphics[width=1\linewidth]{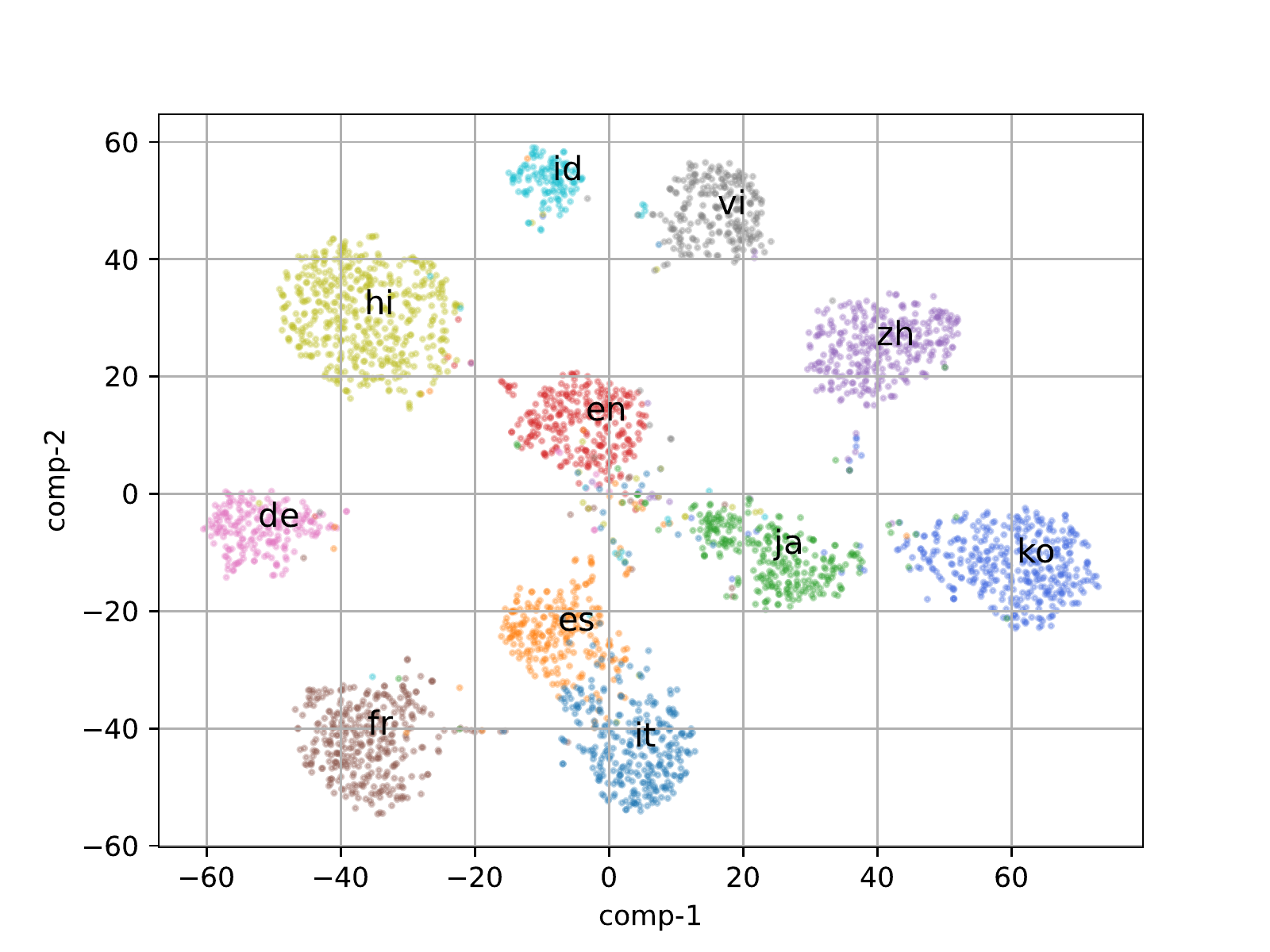} 
\end{subfigure}
\begin{subfigure}{0.45\linewidth}
  \includegraphics[width=1\linewidth]{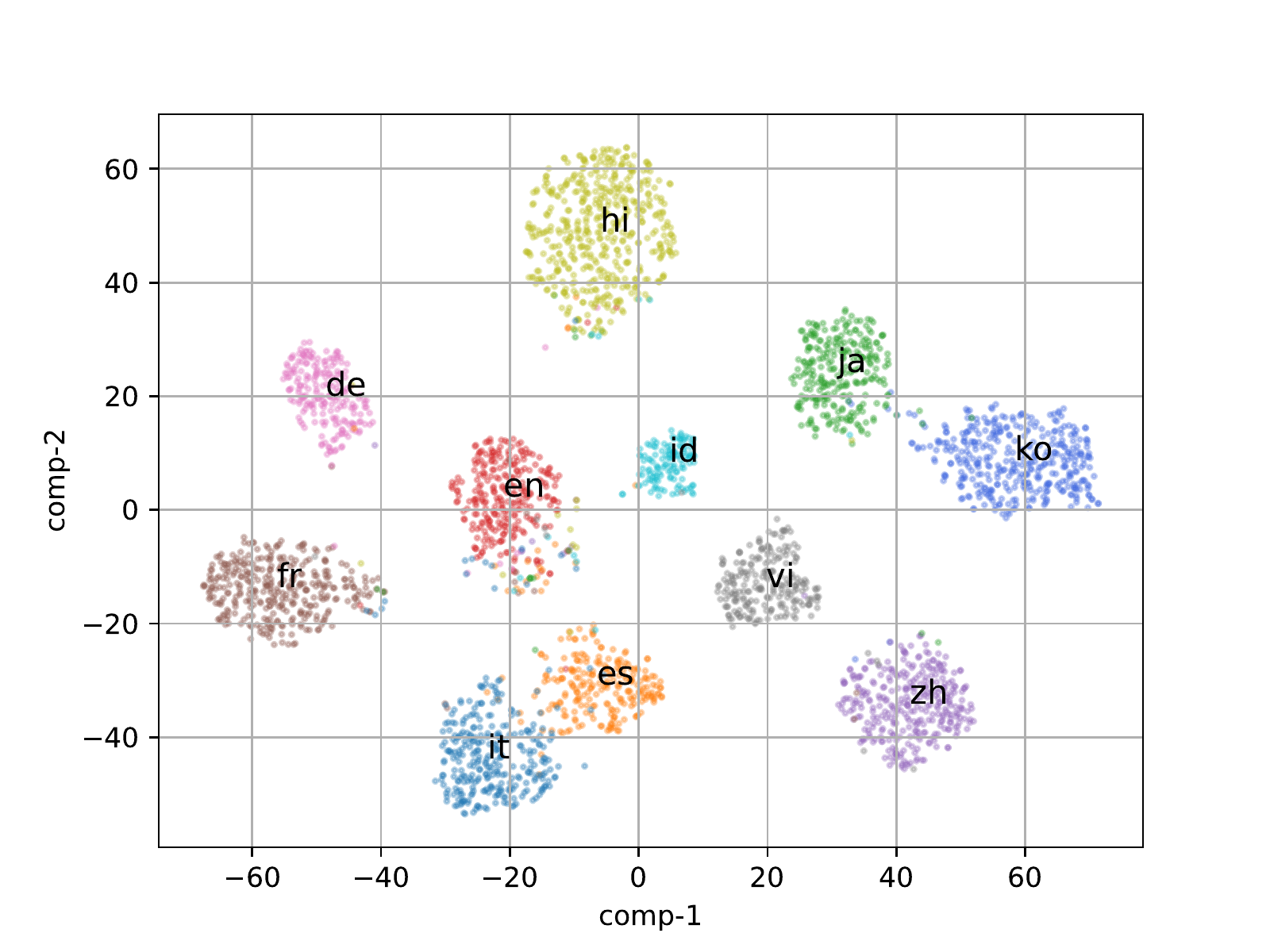} 
\end{subfigure}
\caption{The $t$-SNE plots of two models, trained by supervised-only (left) and joint learning (right). Spanish (es) and Italian (it) were separated more clearly in the joint learning model than in the supervised-only model.}
\label{fig-tsne}
\end{figure*}

\subsection{Performance on VoxLingua107-T11}
We used the VoxLingua107~\cite{valk2021voxlingua107} dataset for the validation. VoxLingua107 contains 107 languages in the training dataset, while the dev set has only 33 languages. In particular, there are no dev sets for hi, id, ko, and vi among the languages we are targeting. The number of utterances for each language in the dev set is also small, with an average of 48 utterances. For this reason, we composed 11 language evaluation datasets (VoxLingua107-T11), collecting sub-three-second utterances from VoxLingua107 training data. The total number of utterances is 3,255, and their average length is 2.54 seconds. 

Table \ref{table-multi-1} shows the classification error rate (\%) according to model structure (top three rows) and learning methods (bottom four rows). First, we checked that the attentive temporal pooling Conformer-based LID (ATP-Conformer)~\cite{wang2022attentive} showed better performance than the ECAPA-TDNN-based LID~\cite{desplanques2020ecapa, das2021hlt, speechbrain}
(even though the ATP-Conformer has a lower number of parameters).
As our base model is slightly larger than the ATP-Conformer, there was a slight performance improvement.

In order to compare the performance of models trained by supervised-only versus joint learning for masking size, we trained models with variable masking sizes from 0 ms to 480 ms (Table \ref{table-multi-1} and Fig. \ref{fig-ablation}).
Both methods showed the best performance when training with a masking size of 240 ms.
Comparing the results of the third and fourth lines of Table 1, it can be seen that even the models trained with the same supervised-only learning show performance differences depending on the presence of masking, which seems to be the effect of SpecAugment~\cite{park2019specaugment}. 
Joint learning once again significantly improved performance. 
As a result, the best error rates for supervised-only and joint learning are 4.94\% and 4.17\%, yielding a relative 15.6\% error rate reduction by the proposed joint learning method. 

Fig. \ref{fig-ablation} shows the LID error rates and pseudo-label classification accuracy for masking size.
Pseudo-label accuracy was 48.43\% when the masking size was 80 ms and decreased to 23.51\% according to the masking size increase to 480 ms.
We found that pseudo-label classification accuracy is not directly related to the LID error rate and that the proposed joint learning method showed a lower LID classification error rate than the best model trained as supervised-only for all masking cases.
The supervised-only models recorded an average error rate of 5.45\% while the joint learning models recorded an average error rate of 4.50\%.

\subsection{Performance for each language}
Table \ref{table-each-lang} shows the performance of two models that were trained by different methods for each language.
Both models performed relatively poorly in Vietnamese, Indonesian, and Spanish. There were many cases where Vietnamese was incorrectly recognized as Indonesian, and Spanish was incorrectly recognized as Italian (or the reverse). 
Comparing the average $F_1$ scores, the joint learning model performed better than the supervised-only model, and there was a significant performance improvement in Indonesian (0.909 $\rightarrow$ 0.931) and Spanish (0.874 $\rightarrow$ 0.912).
Fig. \ref{fig-tsne} shows the $t$-SNE plots of the two models trained by the different methods, and we can see that Spanish and Italian were separated more clearly in the joint learning model.

\begin{table}[t]
\caption{$F_1$ score, precision ($P$), and recall ($R$) of the models as trained by two methods for each language.}
\begin{center}
\begin{tabular}{c|c|cc|c|cc}
\Xhline{2\arrayrulewidth}
\multirow{2}{*}{Lang} & \multicolumn{3}{c|}{Supervised-only best} & \multicolumn{3}{c}{Joint learning best} \\ 
        & $F_1$  & $P$ & $R$        & $F_1$  & $P$   & $R$ \\ \hline
en    & 0.948 &0.988 & 0.912            & \textbf{0.954} &0.988 & 0.923 \\
fr    & \textbf{0.969} &0.965 & 0.974   & 0.968 &0.973 & 0.962 \\
de    & \textbf{0.967} &0.967 & 0.967   & 0.961 &0.937 & 0.985 \\

hi    & 0.961 &0.945 & 0.977            & \textbf{0.976} &0.957 & 0.995 \\
id    & 0.909 &0.887 & 0.933            & \textbf{0.931} &0.902 & 0.962 \\
it    & 0.944 &0.947 & 0.941            & \textbf{0.953} &0.940 & 0.967 \\

ja    & \textbf{0.966} &0.965 & 0.968   & 0.962 &0.974 & 0.950 \\
ko    & 0.981 &0.994 & 0.968            & \textbf{0.983} &0.992 & 0.974 \\
zh    & \textbf{0.953} &0.920 & 0.987   & 0.952 &0.911 & 0.996 \\

es    & 0.874 &0.876 & 0.872            & \textbf{0.912} &0.956 & 0.872 \\
vi    & 0.923 &0.962 & 0.887            & \textbf{0.936} &0.985 & 0.892 \\ \hline

Avg   & 0.945 &0.947 & 0.944            & \textbf{0.953} &0.956 & 0.953 \\ 
\Xhline{2\arrayrulewidth}
\end{tabular}
\end{center}
\label{table-each-lang}
\end{table}

\section{Conclusion}
We proposed context-aware language identification by jointly unsupervised and supervised learning. 
In order to show the effect of our proposed joint learning approach on the wild datasets, 
we collected 11,000 hours of speech data for 11 languages from YouTube for training and used a subset of the VoxLingua107 with sub-three-second utterances. In experiments, the proposed joint learning reduced the error rate by 15.6\% compared to the same structure model trained by supervised-only learning with the same training data.

\bibliographystyle{IEEEbib}
%\bibliography{strings,refs}

\end{document}